\documentclass[jidm,a4paper]{jidm} 
\usepackage{graphicx,url}  
\usepackage[T1]{fontenc}   

\newdef{definition}[theorem]{Definition}
\newdef{remark}[theorem]{Remark}

\usepackage{booktabs}
\usepackage{multirow}
\usepackage[shortlabels]{enumitem}
\usepackage{graphicx} 
\graphicspath{{figures/}} 

\usepackage[table, xcdraw]{xcolor} 
\usepackage{float} 

\usepackage{multirow}
\usepackage[normalem]{ulem}
\useunder{\uline}{\ul}{}
\usepackage{lscape}
\usepackage{booktabs}

\jidmYear{21}
\jidmMonth{August}
\title{Data Modeling for Connected Data \\ A systematic literature review}

\author{Veronica dos Santos\inst{1}}

\institute{Departamento de Informática\\
  Pontifícia Universidade Católica do Rio de Janeiro, Rio de Janeiro, Brazil \\ \email{\{vdsantos\}@inf.puc-rio.br}
}

\begin{abstract}
  A data model specifies how real-world entities and their relationships are represented and operated. In the NoSQL world data modeling usually begins from identifying application queries and designing the data model to efficiently answer them so each database is designed to meet requirements of just one or more applications. But this practice causes a strong coupling between the data model and application queries and promotes data silos. Newly developed applications that manipulate connected data, usually stored in NoSQL Graph Databases, suffer from this type of problem, which is a challenge for data integration projects in Big Data scenarios. This systematic literature review (SLR) was carried out to identify the known approaches for data modeling of connected data. The main contribution of this SLR is an analysis of sixteen works, from 2013 to 2020, in terms of three dimensions: type of contribution, bibliometrics, and data modeling characteristics. Through this analysis, it was possible to identify that reverse engineering of connected data is still a research opportunity since few and incomplete works were found.

\end{abstract}

\category{Information systems}{Data management systems}{Database design and models}[Graph-based database models]
\category{General and reference}{Document types}{Surveys and overviews}


\begin{document}

\begin{bottomstuff}
\end{bottomstuff}

\maketitle

\section{Introduction}

A data model specifies how real-world entities and their relationships are represented and operated \cite{davoudian2018survey}. %
It is specified using a formal modeling language, i.e., any graphical or textual computer language (artificial language) that provisions the design and construction of structures and schemas based on a systematic set of rules.  %
ER, EER, UML, ORM, and ORM2 are examples of data modeling languages. %
Data models of three different abstraction levels can be created (Conceptual, Logical, and Physical) through two approaches: forward and reverse. %

The forward engineering approach starts with data models with higher level, since they are  platform-independent and  describe  a  business  view  of  the  domain  of interest. %
On the other hand, in a reverse engineering approach, the starting point is to extract the physical schema from the database and derive the logical and conceptual models, going in the opposite direction. %
The first approach is generally applied in new applications development while the second is usefull on legacy application maintenance and also for database reuse, integration and migration. %
Abstraction levels enable more general concepts to be represented on a higher level, while the details are gradually introduced on the lower levels. %
The Relational model, proposed by Codd in 1970 \cite{Codd70}, is the most popular data model and successfully adopted the separation-of-concerns of abstraction levels. %

The relational data model caused the emergence of Relational Database Management Systems (RDBMS) and the definition of the Structured Query Language (SQL) as its standard language.  %
Applications from SQL world are schemafull since the schema is defined before data load, that is \textit{schema on-write}.  %
But they were not completely adequate to support all types of applications and new data requirements. %

Newly developed applications, most of them from Big Data scenario, need to incorporate new data aspects. %
Big Data characteristics such as Volume, Variety, Velocity, and so on, motivated the development of data storage and management technologies, such as NoSQL databases and distributed file systems. %
NoSQL databases can be categorized based on their data models into key-value, wide column, document, and graph \cite{davoudian2018survey}. %

One crucial difference between SQL and NoSQL is schema flexibility. %
NoSQL databases are characterized by the lack of a unique fixed and rigid schema, i.e., they are known as schemaless, although some of them support a schema definition.  %
But data always have some structure encoded by the application logic, that is \textit{schema on-read} \cite{davoudian2018survey}.  %
The term  \textit{schema on-read} can be not so accurate since by the time you store the data there is already a schema. %
In the NoSQL world data modeling usually begins from identifying application queries and designing the data model to efficiently answer them so each database is designed to meet requirements of just one or more applications. %
But this practice can cause a strong coupling between the data model and application queries and promotes data silos. %

Connected Data can be defined as data whose interpretation and value requires a good understanding of the relationships between theirs elements \cite{Gudivada2014}. %
Applications that manipulate connected data, such as Social Networks, interaction of proteins in biological systems, Data provenance and Knowledge Graphs, consider entities' inter-connectivity or topology more important, or as important, as the entity's attributes. %
These applications usually model their datasets as graphs using nodes, edges, and their properties. %
It is also observed variations on this basic definition, including, for example, directed or undirected graphs, labeled or unlabeled edges and nodes, simple or multigraph, hypergraphs, and hypernodes. %
Hypernode is a directed graph that allow nesting of graphs since their nodes (hypernodes) can also be graphs. %
Hypergraph is a generalization of graphs that allow edges to be related to more than two nodes (hyperedge) \cite{AnglesG08}\cite{Angles12}. %

Labeled Property Graph (LPG) and Resource Description Framework (RDF) are the two most commom graph data models used by NoSQL databases such as Neo4J\footnote{\url{https://neo4j.com/}} and AllegroGraph\footnote{\url{https://allegrograph.com/}}, respectively. %
There are also NoSQL graph database systems like GraphDB\footnote{\url{https://graphdb.ontotext.com/}}, which can store both graph data models. %
And Multi Model NoSQL databases, called Polystores, like OrientDB\footnote{\url{https://orientdb.org/}} and Virtuoso\footnote{\url{https://virtuoso.openlinksw.com/}}, which can store one or both graph data models besides other NoSQL data models. %
Neo4J, JanusGraph\footnote{\url{https://docs.janusgraph.org/}} and TigerGraph \footnote{\url{https://www.tigergraph.com/}} are examples of NoSQL Graph Databases that supports schema definition. %
Differently from other types of NoSQL, Graph databases support graph data natively by implementing optimized storage and retrieval features, besides access control and authorization subsystems. %
Compared to relational databases, graph databases are suitable for finding relationships within high volume of data faster since it provides index-free adjacency \cite{Kaur2013}.  %

Based on the previous contextualization, a research project in the Big Data scenario that deals with connected data can start with the research question: %

\textit{What are the known approaches for data modeling of connected data?} %

The objective of this SLR is to answer the research question above. %
A review of the primary studies available related to the topic data modeling of connected data in the context of NoSQL graph databases and Big Data aim to integrate and synthesize evidence to support the development of any novel approach. %
A literature review is able to show how a research proposal contributes in a context since new discoveries are built upon previous ones. %
The contribution of the SLR is to present an analysis of the selected works in terms of three dimensions: type of contribution, bibliometrics, and data modeling characteristics. %

But, before planning a SLR, it is necessary to check the existence of other SLRs or similar studies about the topic of interest. %
Four studies similiar to this SLR were found: \cite{Ribeiro2015}, \cite{Sousa2016}, \cite{Zafar2017}, and \cite{Martinez-Mosquera2020}.  %
Different from all of them, this SLR focused on data modeling strategies applicable only to connected data, NoSQL Graph Databases and NoSQL Polystores that support graph data.  %
Martinez-Mosquera et al. \cite{Martinez-Mosquera2020} did not cover reverse engineering approaches, and this SLR did.
And more broadly than \cite{Zafar2017}, that focused on model-driven engineering, this SLR covered model-driven, ontology-driven and data-driven approaches.  %

The remainder of this paper is organized as follows. %
Next section, Systematic Literature Review, presents the review protocol, its execution and search results, including the Quality Assessment and Data Collection. %
In the third section, Answering the Research Question, the selected works are analyzed and briefly described based on the research question that guided this SLR. %
The last section, Conclusion, indicates threats to validity of this SLR and research opportunities. %

\section{Systematic Literature Review}

In the next sub-sections, SLR planning and execution will be detailed.  %

\subsection{Protocol Definition}

The review protocol establishes the context of the research defining search process, including keywords and sources, criteria for selection and rejection of proposals, quality assessment for the selected works and guidelines for data extraction and synthesizing. %

For this review three digital libraries (DL) were chosen: Scopus, Web of Science (WoS), and ACM. %
These DL contain authentic journals and conferences, offer metadata advanced search options, and enable search results bulk download in standardized formats. %
Two lists of journals and Conferences, from Database System, Software Engineering and Big Data areas, with good reputation and indexed by the selected DLs were also predefined with the advisory of a senior researcher. %
A keywords list was created as (“Big Data” OR “Bigdata” OR “Big-data” OR "NoSQL") AND (“Data modeling” OR “Data model” OR “Database design” OR “model-driven” OR “modelagem de dados” OR “modelo de dados” OR “projeto de banco de dados”) combining keywords in English and Portuguese to search for papers indexed by these DLs. %
It is important to emphasize that, as the research question is inserted in the context of Big Data, the expression (with its variations) was included in the list of keywords to restrict the result to the modeling approaches applicable to it. %

Finally, inclusion criteria of the research publications are based on five parameters: %

\noindent 
\begin{enumerate}[label=\alph*]
\item Papers that mention the selected keywords in Title, Abstract or Keywords %
\item Primary research articles from the selected Conferences and Journals %
\item Article, Proceedings Paper, Book Series %
\item Papers in English or Portuguese %
\item Studies based or focused on Data Modeling of NoSQL graph databases exclusively or Multi Model NoSQL databases, including graph data model %
\end{enumerate}

A dataset \cite{ZenodoDS} published at Zenodo contains three files with details of SLR parameters: the list of selected journals (\textit{Selected Journals ranked by JCR.csv}) and selected conferences (\textit{Selected Conferences ranked by h5-index.csv}), besides the DL queries (\textit{Keyword-based queries by Digital Library.csv}). %

\subsection{Search and Selection}

\begin{figure}[ht]
\centering
\includegraphics[scale=.55]{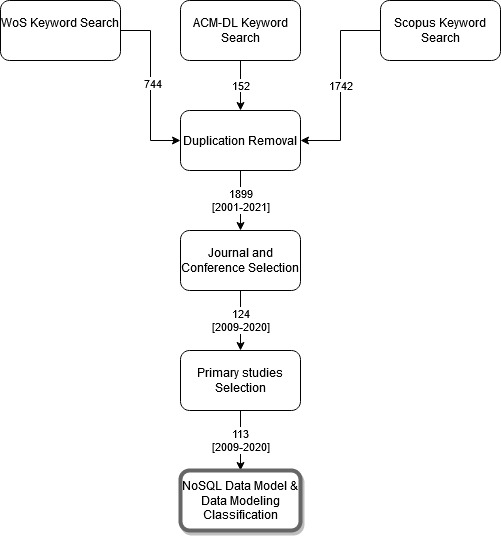}
\caption{SLR execution steps and results}
\label{fig:slr-exec}
\end{figure}

The review search was executed on September 28th, 2020. %
The overview of this phase is illustrated in Figure \ref{fig:slr-exec}. %
The results were downloaded in BibTeX format and Mendeley Reference Manager was used to organize, remove duplicates and filter them. %
Bibliography metadata from all the retrieved works is available in file \textit{SLR 28092020.bib} \cite{ZenodoDS}.  %

Inclusion criteria (a), (b), (c), (d) were applied on the initial 1899 papers based on Title, Abstract, Keywords, and Publication and 113 papers remained. %
Table \ref{tab:incl-excl-criteria} from appendix details the results. %
Finally, a Classification step was carried out in order to identify the studies covering NoSQL Graph Databases and satisfy inclusion criteria (e). %
It is important to mention that some multi model works, although claiming to be system-independent design methodologies for NoSQL databases in general, did not cover graph databases, since they are not aggregate-oriented \cite{NoSQLDistilled2012}. %
During this step, the analysis was extended to reading the introduction and conclusion sections of 27 studies since the Abstract of these works were not clear about graph databases coverage.  %

Ninety seven works were excluded. %
Most of them (71 articles) do not cover data modeling or NoSQL. %
A few of them are proposals that deal with XML (eXtensible Markup Language) data and, although XML can be treated as an ordered-tree-like structure or rooted directed connected graphs, which is a restricted type of graph, in these works the data is modeled and manipulated as documents. %
Eleven of the excluded work are NOSQL multi model proposals that do not include graphs such as NoAM \cite{Bugiotti2014} and NoSE \cite{Mior2017}. 
The selection phase was concluded with sixteen papers from seven journals and conferences. %
Table \ref{tab:selected-works} present the final sixteen selected works separated by two groups and publication year. %

\begin{center} 	
\begin{table}[ht]  \centering
\caption{Selected Works by Type and Year}
\sffamily 
\begin{tabular}{|>{\centering}p{14em}|>{\centering}p{14em}|l|}
\hline
\textbf{Graph}    & \textbf{Multi Model}       & \textbf{Year} \\ 
\hline
-                 & {\cite{Kaur2013}}                   & 2013          \\ \hline
{\cite{DeVirgilio2014}} {\cite{Souza2014}} & -                          & 2014          \\ \hline
{\cite{Neumayr2015}}          & {\cite{Priebe2015}}                   & 2015          \\ \hline
{\cite{Marcin2016}} {\cite{Daniel2016}} & {\cite{DeFreitas2016}} {\cite{Banerjee2016}}          & 2016          \\ \hline
{\cite{Comyn-Wattiau2017}} {\cite{Akoka2017}} & {\cite{Abdelhedi2017}}                   & 2017          \\ \hline
-                 & {\cite{Bjeladinovic20181}} {\cite{Pulgatti2018}} {\cite{Zdepski2018}} & 2018          \\ \hline
-                 & {\cite{Abdelhedi2020}}                  & 2020          \\ \hline
\end{tabular}
\rmfamily 
\label{tab:selected-works}
\end{table}
\end{center}

\subsection{Quality Assessment}

In order to evaluate the importance of the outcomes of the selected works related to the research question, a quality assessment was defined based on seven questions listed below. %
Each work received a punctuation as presented in table \ref{tab:qual-assess}, with maximum of 11 points. %

\noindent 
\begin{enumerate}
 \item Does the study cover NoSQL Graph exclusively?  (1=Yes, 0=No)
 \item Does the study cover reverse engineering? (1=Yes, 0=No)
 \item How many data abstraction levels does the study cover in data modeling ? (0..3)
 \item Are the systems/tools/datasets used/created in the study publicly available? (2=Yes, 1=Partially, 0=No)
 \item Does the study contain information about the experiment design and experiment results?  (2=Yes, both; 1=Yes, only one; 0=No)
 \item Does the study clearly state the limitations of the approach?  (1=Yes, 0=No)
 \item Has the study compared its experiment results with related ones?   (1=Yes, 0=No)
\end{enumerate}

After this step it was possible to identify that: %
\noindent 
\begin{itemize}
\item Only four works propose reverse engineering approaches. %
\item Ten works cover the three abstraction levels and four cover two levels. %
\item Seven have described both the experiment design and results but eight don’t provide any information about experimental evaluation; they used a case study or even an example to illustrate their proposals. %
\item Only two studies compared their experiment results with related ones. %
\item Half of the studies clearly stated their limitations. %
\end{itemize}

\begin{table}[ht] \centering
\caption{Quality Assessment Punctuation}
\sffamily 
\begin{tabular}{|l|r|r|r|r|r|r|r|r|}
\hline
\multicolumn{1}{|c|}{} &
  \multicolumn{1}{c|}{Q1} &
  \multicolumn{1}{c|}{Q2} &
  \multicolumn{1}{c|}{Q3} &
  \multicolumn{1}{c|}{Q4} &
  \multicolumn{1}{c|}{Q5} &
  \multicolumn{1}{c|}{Q6} &
  \multicolumn{1}{c|}{Q7} &
  \multicolumn{1}{c|}{Total} \\ \hline
{\cite{DeVirgilio2014}}  & 1 & 0 & 3 & 1 & 2 & 1 & 1 & 9 \\ \hline
{\cite{Comyn-Wattiau2017}}  & 1 & 1 & 3 & 1 & 1 & 1 & 0 & 8 \\ \hline
{\cite{Neumayr2015}}  & 1 & 0 & 3 & 2 & 0 & 1 & 0 & 7 \\ \hline
{\cite{Marcin2016}}  & 1 & 0 & 1 & 1 & 2 & 1 & 1 & 7 \\ \hline
{\cite{Daniel2016}}  & 1 & 0 & 3 & 2 & 0 & 1 & 0 & 7 \\ \hline
{\cite{DeFreitas2016}}  & 0 & 1 & 3 & 1 & 2 & 0 & 0 & 7 \\ \hline
{\cite{Abdelhedi2020}} & 0 & 1 & 2 & 1 & 2 & 1 & 0 & 7 \\ \hline
{\cite{Abdelhedi2017}}  & 0 & 0 & 3 & 1 & 2 & 0 & 0 & 6 \\ \hline
{\cite{Bjeladinovic20181}}  & 0 & 1 & 0 & 1 & 2 & 0 & 0 & 6 \\ \hline
{\cite{Pulgatti2018}}  & 0 & 0 & 2 & 2 & 2 & 0 & 0 & 6 \\ \hline
{\cite{Akoka2017}}  & 1 & 0 & 3 & 1 & 0 & 0 & 0 & 5 \\ \hline
{\cite{Kaur2013}}  & 0 & 0 & 3 & 2 & 0 & 0 & 0 & 5 \\ \hline
{\cite{Priebe2015}}  & 0 & 0 & 3 & 1 & 0 & 1 & 0 & 5 \\ \hline
{\cite{Souza2014}}  & 1 & 0 & 2 & 0 & 0 & 1 & 0 & 4 \\ \hline
{\cite{Banerjee2016}}  & 0 & 0 & 2 & 1 & 0 & 0 & 0 & 3 \\ \hline
{\cite{Zdepski2018}}  & 0 & 0 & 3 & 0 & 0 & 0 & 0 & 3 \\ \hline
\end{tabular}
\rmfamily 
\label{tab:qual-assess}
\end{table}

\subsection{Data Collection}

In this section, bibliometrics, data modeling characteristics, and a brief description will be presented to better describe the selected works. %

\subsubsection{Bibliometrics}

Bibliometrics is used to describe patterns of publication within a given research topic. %
For this SLR, data collection covered authors, their affiliations, impact factor, number of publications, and country besides paper publication year, publisher, and citation count of the selected works. %
All bibliometric data collected is detailed in file \textit{Bibliometrics Data.pdf} at Zenodo\cite{ZenodoDS}. %

As shown in table \ref{tab:selected-source}, the sixteen selected works range from 2013 to 2020. %
Seven of them focus in graph model, four of which are from ER Conference. %
Nine are multi model that include graph, five of which are from ICEIS Conference. %
IEEE International Conference on Big Data has works from both groups. %
The most cited study is the older one Multi Model \cite{Kaur2013} from 2013 with 43 citations, the second is one Graph study \cite{Marcin2016} from 2016 with 27 citations. %
Both citations counts mentioned are from IEEExplorer but this DL does not have this information for all the selected works. %

\begin{table}[h]
\caption{Selected Works by Type and Publisher}
\begin{tabular}{|>{\centering}p{10em}|>{\centering}p{10em}|r|p{20em}|}
\hline
\textbf{Graph} & \textbf{Multi Model} & \textbf{Quantity} & \textbf{Published in} \\ \hline
- & {\cite{DeFreitas2016}} {\cite{Abdelhedi2017}} {\cite{Pulgatti2018}} {\cite{Zdepski2018}} {\cite{Abdelhedi2020}} & 5 & International Conference on Enterprise Information Systems (ICEIS) \\ \hline
{\cite{Neumayr2015}} {\cite{Akoka2017}} {\cite{Daniel2016}} {\cite{DeVirgilio2014}} & - & 4 & International Conference on Conceptual Modeling (ER) - Lecture Notes in Computer Science \\ \hline
{\cite{Comyn-Wattiau2017}} & {\cite{Kaur2013}} {\cite{Priebe2015}} & 3 & IEEE International Conference on Big Data (IEEE Big Data) \\ \hline
{\cite{Marcin2016}} & - & 1 & IEEE Transaction on Knowledge and Data Engineering (TKDE) \\ \hline
- & {\cite{Bjeladinovic20181}} & 1 & Enterprise Information Systems (EIS) \\ \hline
- & {\cite{Banerjee2016}} & 1 & International Journal of Software Engineering and its Applications \\ \hline
{\cite{Souza2014}} & - & 1 & International Conference on Information Integration and Web-based Applications \& Services (iiWAS) \\ \hline
\end{tabular}
\label{tab:selected-source}
\end{table}

There is a total of 36 research authors from eight different countries and 20 different institutions. %
France is the country with more works (5), followed by Brazil (4), although both have the same number of authors and institutions involved. %
Two papers from Graph group, \cite{Comyn-Wattiau2017} and \cite{Akoka2017}, have two co-authors and other two papers from Multi Model group, \cite{Abdelhedi2017} and \cite{Abdelhedi2020}, have four co-authors. %
None of the authors has papers from different groups. %

\subsubsection{Data Modeling Characteristics}

For each selected work, the following data modeling characteristics were extracted while reading the entire paper: %

\noindent 
\begin{itemize}
 \item Which Big Data V-Dimensions does the study explicitly consider (e.g. velocity, veracity, variety, or others) ? 
 \item In case of Multi Model, which other data models (besides graph) does the study cover (e.g. Column, Document, Key-value, Relational, or others) ?
 \item Which Graph Data Model does the study cover (RDF, LPG, Any,  or others) ?
 \item Which of the three data abstraction levels does the study cover?  
 \item What models did the study propose for each modeling abstraction level? 
  \item Which modeling languages were used? 
  \item Which techniques were applied to transformations between abstraction levels?
  \item Which modeling methodologies (e.g. data-driven, domain-driven) were used? 
  \item Which data model generation approach (e.g. forward, reverse, others) were used? 
  \item In case of reverse engineering, what sources were used to extract the schema?
\end{itemize}

Tables \ref{tab:dm1} and \ref{tab:dm2} from \ref{sec:apendix} contain the details of each selected work for the characteristics mentioned above. %

Neo4J and LPG model was used in half of the studies. %
None Multi model proposals dealt with RDF model although Relational Model, that is not a NoSQL data model, appeared in five of the nine studies. %
Jena TDB\footnote{\url{https://jena.apache.org/documentation/tdb/index.html}}, a native graph store from open-source Jena framework, was used by \cite{Neumayr2015}. %
Diplocloud is a new native RDF Storage Model proposed by \cite{Marcin2016} but the authors did not mention whether it is freely available on the Web for download and test. %
UMLtoGraphDB \cite{Daniel2016}, as well as the proposal of \cite{DeVirgilio2014}, used Blueprint TinkerPop API\footnote{\url{https://github.com/tinkerpop/blueprints}} as an abstraction layer at physical level. %

In terms of modeling languages distribution, we identified that eleven different languages were used. %
Peter Chen's notation (ER) \cite{PeterChen76} and its variation Enhanced Entity Relationship (EER) are the predominant for the Conceptual Level, used in eight works, followed by UML. %
According to \cite{AnglesG08}, semantic database models such as ER are able to express the graph-like structure generated by the relationships among the modeled entities. %
A node corresponds to an entity, its attributes are modeled as node properties and relationship between entities translates into edges between nodes. %

Despite that it is still used in NoSQL world. %
OCL (Object Constraint Language) was used only in one work \cite{Daniel2016} for Model-to-Model transformation to generate database-level queries from business rules and invariants defined using OCL. %
Ontology Web Language (OWL), was used exclusively in the two ontology-driven works. %

Three methodologies were identified among the selected works: data-driven, model-driven and ontology-driven. %
Model-driven is the predominant methodology among Graph works (five from seven) and data-driven among Multi Model works (five from nine). %
The majority of the works (nine) followed the forward engineering approach. %
Reverse engineering was covered exclusively only by two works. %
Two Multi Model works, covering migration and interoperability, were also considered in this review since they deal with mappings of different database data models within the same abstraction level or between different abstraction levels. %

\subsubsection{Brief Description}

\begin{figure}[ht]
\centering
\includegraphics[scale=.48]{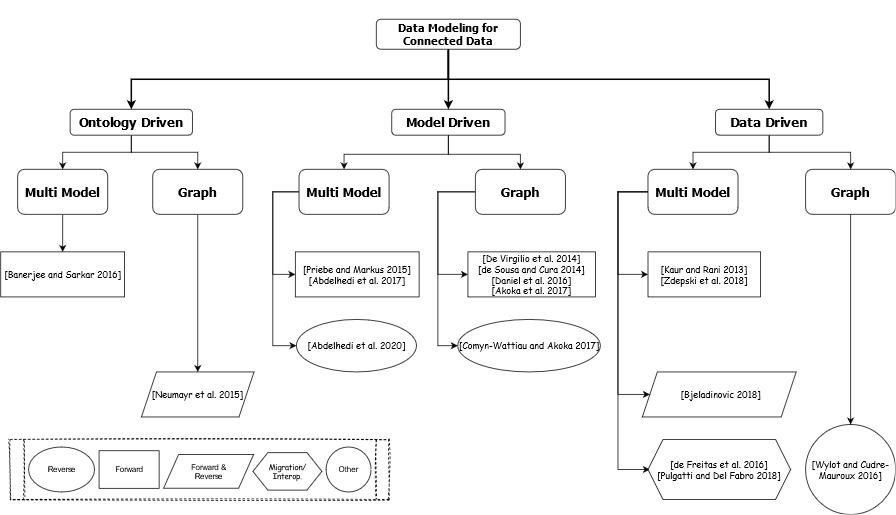}
\caption{Selected Works classification by methodology, group, and approach}
\label{fig:SLR_Taxonomy}
\end{figure}

In this section there is a brief description of each work. %
The selected papers are grouped by methodology, as seen in figure \ref{fig:SLR_Taxonomy}, and presented chronologically. %
The description highlights the main characteristics of each work, including their limitations and strengths.  %

\textbf{Ontology-Driven}

Neumayr, et al. \cite{Neumayr2015}, in 2015, proposed a guideline to aggregation data modeling and aggregation view generation. %
One named RDF graphs is created for each Aggregated View based on the aggregation level. %
This work is the only that focused in Online Analytical Processing (OLAP). %

A Domain Ontology and an Aggregation Ontology are used in the conceptual abstraction level and represented in OWL. %
At the logical level, a Named RDF data model is created and stored in Jena TDB. %
The transformation between levels is based on RDFS reasoning and three SPARQL generic updates. %
This work used both forward and reverse approaches and has a section dedicated to explain its limitations and propose extensions. %

The other work of this methodology, \textit{Ontology Driven Meta-Modeling for NoSQL Databases: A Conceptual Perspective} \cite{Banerjee2016}, was published in \textit{International Journal of Software Engineering and Its Applications} and belongs to the Multi Model group. %
An ontology driven meta-model, called Ontology Driven NoSQL Data Model (ODNSDM), was proposed to conceptualize data representation facets over heterogeneous kinds of databases (SQL and NoSQL). %

ODNSDM can be realized as a layered organization composed of three main layers – Collection, Family and Attribute. %
In addition, ODNSDM is composed of a formally expressed axiom set: six for relationships, six for collections, nine for family, five for attributes, and ten for properties. %
Unfortunately, an OWL file representing ODNSDM is not public available. %

\noindent
\textbf{Model-Driven}

Model-driven design (MDD) is a development paradigm strongly based on models. %
Models are considered the primary artifacts of the development process. %
A model-driven approach (MDA) relies on the use of Object Management Group (OMG) standards to materialize MDD. %
These standards aim to enable automated or semi-automated model transformations and code generation. %
MDA comprises three model levels: Computation Independent Model (CIM), Platform Independent Model (PIM), and Platform Specific Model (PSM). %
And the transformation between levels is defined by query/view/transformation (QVT) rules. %
ATL is a domain specific language for defining model-to-model (M2M) transformations aligned with the QVT standard. %
ATL provides both declarative (rule-based) and imperative constructs for transforming and manipulating models. %

The older model-driven study from Graph group was published in ER 2014 by De Virgilio et al. \cite{DeVirgilio2014} and named \textit{Model-Driven Design of Graph Databases}. %
It described a forward engineering design process based on an automatic analysis of a conceptual representation of the domain of interest formalized using any conceptual data model. %
Although the authors illustrated using an ERD, they claimed that it can be easily adapted to use a UML diagram as input. %
The process is organized in three different phases: (i) generation of an oriented ER diagram, (ii) partitioning of the elements (entities and relationships) of the obtained diagram and (iii) definition of a template over the resulting partition. %
Two strategies were used: compact (merging different nodes) and sparse (the properties of an object with n properties is decomposed into a set of n different nodes) strategies. %

First, they transform an ER diagram, which is an undirected and
labelled graph, into a directed, labelled and weighted graph, called Oriented ER (O-ER) diagram using 3 rules. %
The partitioning of the O-ER diagram is based on other 3 rules for grouping together elements of the O-ER diagram so that every element of the diagram belongs to one and only one group. %
The logical schema is the template generated to describe the data types occurring in a graph database and the ways they are connected. %
The authors claimed that this schema can be converted into any NoSQL GraphDB and stated that the proposal does not consider transaction requirements and query operation loads and the methodology needs improvements to disambiguate between possible decisions. %

In 2016, Daniel et al. \cite{Daniel2016} proposed \textit{UMLtographDB: Mapping conceptual schemas to graph databases}. %
A forward engineering process with three transformation steps supported of two meta-models and 110 transformation rules to create a middleware, composed of a set of Java Classes, wrapping the structure of the database. %
A prototype is avaliable at GitHub \footnote{\url{https://github.com/atlanmod/UML2NoSQL/}}. %
At the CIM level there are two models: UML Class Diagram and 
OCL Constraints. %
GraphDB metamodel is a canonic graph representation used to define the PIM model. %
Blueprints API and Gremlin were used at the PSM level. %

UMLtoGraphDB framework translates
conceptual schemas expressed using UML into GraphDB model that conforms to a metamodel using a model-to-model (M2M) transformation, named Class2GraphDB, built with the ATL language (aligned with QVT standard). %
Another M2M, OCL2Gremlin, generates database-level queries  from business rules and invariants defined using the Object Constraint Language (OCL). %
The last step is Graph2Code, a PSM-to-code transformation that creates the middleware. %
As a limitation, the authors warned that GraphDB has no construct to explicitly represent inheritance so the mapping creates duplicate elements to facilitate queries. %

Comyn-Wattiau and Akoka \cite{Comyn-Wattiau2017} proposed \textit{Model driven reverse engineering of NoSQL property graph databases: The case of Neo4j}, and both authors also proposed \textit{A Four V’s Design Approach of NoSQL Graph Databases} \cite{Akoka2017} in conjunction with Prat. %
These two works were published in 2017. %

A reverse engineering process with four steps supported by two meta-models and transformation rules is presented in \cite{Comyn-Wattiau2017}. %
The first transformation rule is responsible to map create statements, at the PSM level, into an instance of GraphDB meta-model, at the PIM level. %
The next transformation (M2M) converts edges and vertices to an instance of EER meta-model at the CIM level. %
Neo4J was used in the illustrative example. %
A prototype implementing the meta-models and the transformation rules was under development at the time of the article publication. %
The limitation of the proposal is that it does not consider hypergraph so the relationships resulting from reverse process are all binary. %

A forward engineering process with four steps supported by two meta-models and transformation rules considering Big Data V-dimensions is the scope of Akoka et al. \cite{Akoka2017}. %
At the CIM level, a V’s EER metamodel is defined to cover data aspects: Volume, Variety, Velocity, and Veracity. %
An EER logical property graph metamodel supports the PIM level. %
And at the PSM level the authors used Neo4J. %

Transformation rules are defined to map elements of V’s EER instance to an instance of logical graph meta-model and then to convert it into a script to generate a Neo4J database. %
Physical database creation considered the volume
requirement, collected at the beginning of the design process, in order to generate instances. %
But the other dimensions were not explored in the illustrative case since they do not propagate to the physical level. %

The authors did not conduct a robust experiment to test their forward engineering approach. %
They presented an illustrative scenario in order to assess the utility and applied it in Neo4j. %
They stated that their approach would benefit from more experiments on large databases and they were also developing a prototype to implement the meta-models and the transformation rules by the time of article publishing. %

Sousa and Cura \cite{Souza2014} proposed, in 2014, a forward engineering design methodology based on four algorithms to convert a Extended Binary Entity-Relationship Model (EB-ER) into a canonic Graph Based Logical Data Model. %
The algorithms are: Main, CardinalityConstraints, VertexConstraints, and EdgeConstraints. %
EB-ER defines four types of attributes and dominant roles but it lacks of ER concepts such as specialization / generalization relationship and  associative / weak entities. %

\textit{Business information modeling: A methodology for data-intensive projects, data science and big data governance}
\cite{Priebe2015} was published in \textit{IEEE International Conference on Big Data} in the year of 2015. %
An organization-wide and unique catalog englobing all relevant user communities’ requirements is used to represent business concepts definitions and data links for data-intensive projects like Data Warehouse implementations and enable data governance. %
It is supported by BIM metamodel implemented in Accurity Glossary\footnote{\url{https://www.accurity.ai/platform/glossary/}}. %
At the conceptual level, besides BIM metamodel, an EER representation is also used. %
Three extensions from ER are considered: (i) attribute definition is uniquely defined attribute name and its description attached to multiple entities, (ii) inheritance on attribute definition level to enable a specialized version of a more general attribute definition, and (iii)  non-atomic (composite) attributes definition. %

BIM considers the variety dimension of Big Data as information items can be both of structured and unstructured. %
So, at physical level, different data models are represented using three general concepts in their information: (i) Data Set, corresponding to database, schema or directory, (ii) Data Structure, corresponding to table, view, file, document, collection, node or relationship, and (iii) Data Element corresponding to column (within a table or view), field or property. %
According to the authors, the application of BIM to big data environments and data science is work in progress. %

Another Multi Model work, presented in ICEIS, is \textit{Logical Unified Modeling for NoSQL Databases} \cite{Abdelhedi2017}. %
Here a forward engineering process named UMLtoNoSQL is described. %
The process is composed by two steps supported by one generic meta-model and transformation rules that transforms a conceptual data model describing Big Data into several NoSQL physical models. %
A UML Class Diagram is used at the CIM level and a model-to-model transformation, named UMLtoGenericModel and based on QVT, converts from CIM to PIM. %
The instance of Generic Logical Model at PIM level is transformed into a PSM by another model-to-model transformation rule: GenericModeltoPhysicalModel. %

In terms of NoSQL database systems, they used Cassandra (Column), MongoDB (Document), and Neo4J (LPG). %
Neo4J physical model  was defined as a tuple (V, E), where V is a set of vertex and E is a set of edges. %
Three rules were described to convert the elements of the generic logical model (table, attributes and relationships) into a NoSQL graph physical
model (vertex, property and edges). %
The authors conducted an experimental assessment of their proposal using: (1) Eclipse Modeling Framework (EMF), (2) Ecore, a metamodeling language, used to create the Source and Target metamodels and (3) XML Metadata Interchange (XMI), an XML based standard for metadata interchange. %

The most recent work is from this year, ICEIS 2020, \textit{Discovering of a conceptual model from a NoSQL database} \cite{Abdelhedi2020}. %
It shares three authors of the previous model-driven multi model study but, different from it, this work is about reverse engineering. %
Their proposal is about a reverse engineering process named ToConceptualModel whose input is a NoSQL physical model (PSM) and the output corresponds to the conceptual model (UML class diagram). %
The model-to-model transformations between PSM, generated by ToPhysicalModel, and PIM are formalized by seven rules expressed in QVT. %
The authors considered their work as a Multi Model, including graph, automated process although the experiment used only NoSQL Document Database. %
Besides this, the Source Metamodel is using elements, like collections, that may not be applicable to graph data models. %
This work does not consider inheritance, data aggregation, and N-ary relationships, the later is a requirement for hypergraphs. %
An experimental evaluation comparing the productivity of developers while writing queries with or without models was carried out to validate the generated model. %

\noindent
\textbf{Data-Driven}

Data-driven modeling is a technique that designs the data models based on how the data are organized within the dataset and how they are derived from external systems. %

The older study from all the selected works is one from Multi Model group \cite{Kaur2013} named \textit{Modeling and querying data in NoSQL databases}, written by K. Kaur and R. Rani and published in \textit{2013 IEEE International Conference on Big Data}. %
It reported a case-study explanation to illustrate forward engineering approach for data modeling in practice using MongoDB and Neo4J. %
An ERD was used at the conceptual level, a UML Class Diagram and Graph Diagram at the Logical level for Document and graph based modeling. %
This article, although being the most cited one, was just describing the models, not discussing or proposing a tool or a method or explained the transformation between levels. %

Wylot and Cudre-Mauroux, in 2016, proposed \textit{DiploCloud: Efficient and Scalable Management of RDF Data in the Cloud} \cite{Marcin2016}. %
This work, from the Graph group, focused only in the physical level so some characteristics of others works like modeling languages, transformation between levels and approach are not applicable. %
DiploCloud is a system design of a cloud-based distributed RDF database system. %
It may be considered tuning instead of data modeling but a Physical Storage Model based on key index, templates, partitioning police, and auxiliary indexes is defined and validated. %
The modeling aims to balance between intra-operator parallelism and data colocation but with the cost of redundancy and more complex inserts and updates (limitations). %

Also in 2016, a work by Brazilian researches named \textit{Conceptual mappings to convert relational into NoSQL databases} \cite{DeFreitas2016} was published in the \textit{International Conference on Enterprise Information Systems}. 
A set of conceptual mappings from Relational databases into any NoSQL database and two conversion algorithms, called R2NoSQL, to support the data conversion process are proposed. %
This work dealt with Relational, Key-Value, Document, and Column data models, besides LPG, at the logical level. %
At the conceptual level a ERD was used and RDBMS, with RDBMS metadata as the source of reverse engineering, at the physical level. %
For the experimental evaluation, they developed a tool prototype, which implements a case study with a Relational database and a Document based NoSQL system, using MySQL and MongoDB. %

Bjeladinovic proposed a process of hybrid SQL/NoSQL database design from scratch, in a forward engineering approach, and a database redesign, reverse and forward engineering approach. %
The article \textit{A fresh approach for hybrid SQL/NoSQL database design based on data structuredness} \cite{Bjeladinovic20181} was published in 2018. %
The process starts with requirements gathering and analysis to decide which technology and design path will be applied. %
Different types of requirements for database design, including non-functional such as performance and scalability, define the technology and the design path (SQL or NoSQL). %

In terms of database systems, SQL is the preferred route, but the process used an indicator of data structuredness that points if a table is a preliminary candidate for switching to a NoSQL database. %
This indicator is calculated based on RDBMS metadata and workload (queries and DDL). %
The author claimed that the process is applicable to any NoSQL data model, including graph, using any conceptual model and any modeling language. %
However, at the logical level and NoSQL path, the process is based on NoAM \cite{Bugiotti2014} and NoAM's authors did not consider graph databases. %

Two Multi Model works were published in ICEIS 2018: \textit{JSON-based interoperability applying the pull-parser programming model} \cite{Pulgatti2018} and \textit{An approach for modeling polyglot persistence} \cite{Zdepski2018}. 

Interoperability is the capacity of sharing resources among different systems. %
The proposal of Pulgatti and Didonet \cite{Pulgatti2018} is a NoSQL interoperability solution based on the JSON format with the aid of a pull-parser programming model for executing a set of rules over a stream of objects. %
The authors defined a set of interoperability rules separated by the category of input data model (Key-Value, Column, Document, and LPG graph model) and illustrated the output of each rule execution. %
In this work, the conceptual abstraction level was not covered and the Relation and RDF data model were not included. %

A forward engineering database design process for Polyglot Persistence, covering Relational, Key-Value, Column, Document, and any graph Data Model is described in \cite{Zdepski2018}. %
The proposal extends the classical database design by adding at the logical design phase the steps required to model subsystems with multiple integrated data models. %
The process starts with an ER conceptual model and the transformation from Conceptual to Logical is based on a segmentation step, called Segmentation Unit Definition, which divides the conceptual model in parts according to main features to be implemented. %
Next, a Consistency Units definition is executed, based on the analysis of consistency data requirements (from ACID to BASE) and, as result, data fragmentation (partitioning) can be identified. %
At last, the target data model of each subsystem is defined based on other non-functional requirements. %

\section{Answering the Research Question}

Recapping the research question: \textit{What are the known approaches for data modeling of connected data?} %
By studying and examining the selected works thoroughly, it was possible to extract the different data modeling characteristics such as methodologies, approaches, modeling languages, data models, abstraction levels. %
The Data Collection was conducted in order to characterize the known approaches in terms of these main characteristics, limitations and its contribution. %
The main contributions of each work are listed in table \ref{tab:contribution}. %

\begin{table}[ht] \centering
\caption{Main Contributions}
\sffamily 
\begin{tabular}{@{}ll@{}}
\toprule

\cite{Kaur2013} &
  Case Study presentation \\ \midrule

\cite{Souza2014} &
  \begin{tabular}[c]{@{}l@{}}Extended Binary Entity-Relationship Model (EB-ER)\\ Transformation Algorithms (four) Conceptual-to-Logical \\ Evaluation (Case Study)\end{tabular} \\ \midrule

\cite{DeVirgilio2014} &
  \begin{tabular}[c]{@{}l@{}}Oriented ER (O-ER) diagram\\ Foward Transformation Rules for ER\\ Partitioning Rules of an O-ER Diagram\\ Evaluation (Experiment)\end{tabular} \\ \midrule

\cite{Priebe2015} &
  \begin{tabular}[c]{@{}l@{}}Organization-wide Model\\  Centralized Information Catalog - Business Glossary (Variety) \\ Evaluation (Case Study)\end{tabular} \\ \midrule

\cite{Neumayr2015} &
  \begin{tabular}[c]{@{}l@{}}A vocabulary for ontology-based RDF Analytics \\ Aggregated Views as named RDF graphs\end{tabular} \\ \midrule

\cite{Marcin2016} &
  RDF-based system architecture (Storage Model) \\ \midrule

\cite{DeFreitas2016} &
  \begin{tabular}[c]{@{}l@{}}Conceptual mappings from SQL to NoSQL\\ Source Table Classification and Conversion Algorithm\\ Evaluation (Experiment)\end{tabular} \\ \midrule

\cite{Daniel2016} &
  \begin{tabular}[c]{@{}l@{}}GraphDB metamodel\\ Foward Transformation Rules (QVT/ATL) for UML and OCL\\ Code Generation using Blueprints API ( plugins on Github)\end{tabular} \\ \midrule

\cite{Banerjee2016} &
  \begin{tabular}[c]{@{}l@{}}Ontology Metamodel\\ Algorithms to Schema-less and Schema-based databases\end{tabular} \\ \midrule

\cite{Abdelhedi2017} &
  \begin{tabular}[c]{@{}l@{}}Metamodels\\ Foward Transformation Rules (QVT)\\ Evaluation (Case Study)\end{tabular} \\ \midrule

\cite{Akoka2017} &
  \begin{tabular}[c]{@{}l@{}}EER Conceptual Metamodel\\ EER logical property graph Metamodel\\ Foward Transformation Rules (QVT/ATL) for ER\end{tabular} \\ \midrule

\cite{Comyn-Wattiau2017} &
  \begin{tabular}[c]{@{}l@{}}EER Conceptual Metamodel\\ GraphDB metamodel\\ Reverse Transformation Rules (QVT/ATL) for ER\end{tabular} \\ \midrule

\cite{Zdepski2018} &
  \begin{tabular}[c]{@{}l@{}}Unified NoSQL methodology (Variety) \\ Evaluation (Experiment) \end{tabular} \\ \midrule

\cite{Pulgatti2018} &
  \begin{tabular}[c]{@{}l@{}}Interoperability Rules from JSON (Variety)\\ Pull-parser programming model for streaming (Velocity)\end{tabular} \\ \midrule

\cite{Bjeladinovic20181} &
  \begin{tabular}[c]{@{}l@{}}Unified SQL and NoSQL methodology\\ Guidelines for database technology selection\end{tabular} \\ \midrule

\cite{Abdelhedi2020} &
  \begin{tabular}[c]{@{}l@{}}Metamodels\\ Reverse Transformation Rules (QVT)\\ Evaluation (Experiment)\end{tabular} \\ \bottomrule
  
\end{tabular}
\rmfamily 
\label{tab:contribution}
\end{table}

Most of the works follow the forward engineering approach. %
But there were two papers describing data models generation through reverse engineering: \cite{Comyn-Wattiau2017} and \cite{Abdelhedi2020}. %
Both proposals are model-driven reverse engineering approaches that used the create statements extracted from Neo4J metadata as the source for reverse engineering. %
A set of instances was also used by \cite{Comyn-Wattiau2017}. %
But it is important to remember that, although some NoSQL databases support schema definition, most of them are schema flexible so these proposals can not be applicable when there isn't a schema definition stored in the NoSQL graph databases. %
The proposal from \cite{Comyn-Wattiau2017} does not considered hypergraph so the relationships resulting from reverse process are all binary. %

Both previously mentioned studies were written by authors that also have model-driven forward engineering approaches proposals within the selected works set: \cite{Akoka2017} and \cite{Abdelhedi2017}. %
Similarly, other forward engineering works found by this review can also be used as bases for any novel reverse engineering proposal and comparison. %
Although it is important to be aware of the limitations of these works. %

\cite{DeVirgilio2014} authors' claimed that their logical schema can be converted into any NoSQL Graph Database but stated that the proposal does not consider transaction requirements and query operation loads and the methodology needs improvements to disambiguate between possible decisions. %
Also, GraphDB from \cite{Daniel2016} has no construct to explicitly represent inheritance so the mapping creates duplicate elements to facilitate queries. %
Extended Binary Entity-Relationship Model (EB-ER) \cite{Souza2014} defines four types of attributes and dominant roles but it lacks of ER concepts such as specialization / generalization relationship and  associative / weak entities. %
The proposal from \cite{Abdelhedi2020} did not considered inheritance, data aggregation, and N-ary relationships, the later is a requirement for hypergraphs. %
\textit{DiploCloud} \cite{Marcin2016} focused only in the physical level as it proposes a system design of a cloud-based distributed RDF database system. %

Among Multi Model works, there are proposals that claim to cover any NoSQL data model and did not properly consider the particularities of graph data models. %
Bjeladinovic proposed a process of hybrid SQL/NoSQL database design and redesign, reverse and forward engineering approaches, applicable to any NoSQL data model. %
However, at the logical level and NoSQL path, the process is based on NoAM \cite{Bugiotti2014} and it does not cover graph databases. %
A set of conceptual mappings from Relational databases into any NoSQL database and two conversion algorithms, called R2NoSQL, to support the data conversion process were proposed in \cite{DeFreitas2016}. %
Specifically for Relational to RDF conversion there already exists W3C Standards: RDB2RDF\footnote{\url{https://www.w3.org/TR/rdb-direct-mapping/}} and R2RML\footnote{\url{https://www.w3.org/TR/r2rml/}}. %

The two most popular graph data models, LPG and RDF, have structural differences that must be considered in the data modeling proposals at logical and physical levels. %
It is possible to interchange data between this two graph data models \cite{Oracle2014}. %
But none of the selected works explored the interoperability or migration between them. %
The two Multi Model works, \cite{DeFreitas2016} and \cite{Pulgatti2018}, covering migration and interoperability, only considered the LPG data model. %

None of the works used SPARQL Inferencing Notation (SPIN)\footnote{\url{https://www.w3.org/Submission/spin-modeling/}} or Shapes Constraint Language (SHACL)\footnote{\url{https://www.w3.org/TR/shacl/}} to model business rules or integrity constraints applicable to RDF data. %
In \cite{Angles12} the author observed that integrity constraints were poorly studied in graph databases and, considering the selected works analyzed, it is still a research opportunity. %

Neo4J, a LPG-based system, were used by the majority of the works at physical level but there are other LPG-based systems that can be chosen for experimental evaluation. %
The Blueprint TinkerPop API, as an abstraction layer on top of a variety of graph databases, is an interesting resource to test a system-independent proposals since other NoSQL Databases, such as JanusGraph and BlazeGraph\footnote{\url{https://blazegraph.com/}}, supports this access method. %
None from the reverse engineering selected works is target to Triplestores (RDF-based systems). %

None of the selected works explored the application code nor workload queries as source for reverse engineering. %
Remembering that queries were already used by relational model reverse engineering \cite{Petit95} as input and for conceptual model refinement and, in NoSQL world, data modeling begins by identifying application queries \cite{Priebe2015}, it seems that workload queries can also be helpful as input for a novel reverse engineering approach and have not been explored by the selected works. %

Investigating the subsequent publications of the most recurrent authors identified in the bibliometrics analysis, it was possible to find a reverse engineering proposal that uses queries \cite{Comyn-WattiauAkoka2019}. %
It consists in a four steps method to transform graph-based queries (MATCH statements in Cypher language) into a EER schema using a rule-based mechanism. %
SPARQL, Cypher and Gremlin are the most used graph query languages among NoSQL Graph databases but there is not a graph language standard as SQL is for RDBMS. %


\section{Conclusion}

This SLR presented an analysis of primary studies available related to the topic data modeling of connected data in the context of NoSQL graph databases and Big Data. %
Sixteen works were selected and separated in two groups: Graph and Multi Model. %
The analysis considered three dimensions: type of contribution, bibliometrics, and data modeling characteristics. %
It also identified some limitations and strengths of the known approaches that can help to enhance the proposals or in the development of novel ones. %

Considering that any systematic literature review is subject to threats to validity, it is important to anticipate and recognize them. %
Only one researcher with knowledge in the area of databases and database design carried out the selection of works. %
Each step of the SLR has been clearly documented and the data collected is public available at Zenodo \cite{ZenodoDS} to allow reproductibility and to mitigate this threat. %
This review can be complemented in the future with some new keywords such as "database schema"  (synonym for "physical data model"), "schema inference", and "schema extraction" (synonyms for "reverse engineering"). %

After this SLR, there are evidences that reverse engineering of NoSQL Graph databases is still a research opportunity since few and incomplete works were found, forward engineering approaches can be used as referential for novel proposals as well application source code, workload queries, and data instances as potential sources for schema extraction. %

\bibliographystyle{jidm}
\bibliography{jidm}

\newpage
\appendix

\section{}
\label{sec:apendix} %

\sffamily 

\begin{table}[h]  \centering
\begin{tabular}{|l|r|l|}
\hline
\multicolumn{3}{|c|}{{\ul \textbf{SEARCH}}} \\ \hline
\textbf{Digital Libraries} &
  \multicolumn{1}{l|}{\textbf{Quantity}} &
  \textbf{Period} \\ \hline
WoS &
  744 &
  2010-2020 \\ \hline
ACM-DL &
  \begin{tabular}[c]{@{}r@{}}(9 + 114 + 29) \\ = 152\end{tabular} &
  2012-2020 \\ \hline
Scopus &
  1714 &
  2001-2021 \\ \hline
\multicolumn{1}{|r|}{\textbf{Total}} &
  2610 &
   \\ \hline
\multicolumn{1}{|r|}{\textbf{Total after duplication removal}} &
  1899 &
  2001-2021 \\ \hline
\multicolumn{3}{|c|}{{\ul \textbf{SELECTION}}} \\ \hline
\textbf{Journals \& Conferences} &
  \multicolumn{1}{l|}{\textbf{Quantity}} &
  \textbf{Period} \\ \hline
IEEE International Conference on Big Data &
  36 &
  2013-2018 \\ \hline
\begin{tabular}[c]{@{}l@{}}International Conference on Conceptual Modeling (ER)\end{tabular} &
  21 &
  2014-2019 \\ \hline
\begin{tabular}[c]{@{}l@{}}International Conference on Enterprise Information Systems (ICEIS)\end{tabular} &
  11 &
  2016-2020 \\ \hline
International Conference on Management of Data (SIGMOD) &
  10 &
  2014; 2016-2019 \\ \hline
IEEE International Conference on Data Engineering (ICDE) &
  9 &
  2014-2019 \\ \hline
\begin{tabular}[c]{@{}l@{}}International Conference on Information Integration and \\ Web-based Applications \& Services (iiWAS)\end{tabular} &
  8 &
  2011-2015 \\ \hline
\begin{tabular}[c]{@{}l@{}}International Conference on Very Large Databases (VLDB)\\ Proceedings of the VLDB Endowment\end{tabular} &
  7 &
  \begin{tabular}[c]{@{}l@{}}2009; 2011; \\ 2013-2016\end{tabular} \\ \hline
Journal of Big Data &
  6 &
  2015; 2017-2019 \\ \hline
IEEE Transactions on Big Data &
  4 &
  2019-2020 \\ \hline
IEEE Transaction on Knowledge and Data Engineering (TKDE) &
  4 &
  \begin{tabular}[c]{@{}l@{}}2016; 2017; \\ 2019; 2020\end{tabular} \\ \hline
\begin{tabular}[c]{@{}l@{}}International Journal of Software Engineering and its \\ Applications (IJSEIA)\end{tabular} &
  2 &
  2015; 2016 \\ \hline
SIGMOD Record &
  2 &
  2010; 2015 \\ \hline
\begin{tabular}[c]{@{}l@{}}International Conference on Model and Data Engineering (MEDI)\end{tabular} &
  2 &
  2018 \\ \hline
ACM Computing Surveys &
  1 &
  2018 \\ \hline
Communications of the ACM &
  1 &
  2011 \\ \hline
VLDB journal &
  0 &
  - \\ \hline
Database for Advances in Information Systems (SIGMIS) &
  0 &
  - \\ \hline
Brazilian Symposium on Software Engineering (SBES) &
  0 &
  - \\ \hline
Brazilian Symposium on Information Systems (SBSI) &
  0 &
  - \\ \hline
\multicolumn{1}{|r|}{\textbf{Total (Conference + Journals)}} &
  124 &
  2009-2020 \\ \hline
\multicolumn{1}{|r|}{(-) secondary studies} &
  6 &
   \\ \hline
\multicolumn{1}{|r|}{(-) other types like tutorial, talks, demo, ...} &
  4 &
   \\ \hline
\multicolumn{1}{|r|}{(-) paper with an extended version of the same work (NoSE)} &
  1 &
   \\ \hline
\multicolumn{1}{|r|}{\textbf{Total (after inclusion/exclusion criteria a,b,c,d)}} &
  113 &
   \\ \hline
\multicolumn{3}{|c|}{{\ul \textbf{CLASSIFICATION}}} \\ \hline
\begin{tabular}[c]{@{}l@{}}Multi Model NoSQL Databases, \\ excluding Graph\end{tabular} &
  11 &
  \begin{tabular}[c]{@{}l@{}}2014{[}2{]}; 2015{[}3{]}; \\ 2016{[}2{]}; 2018{[}2{]}; 2019{[}2{]}\end{tabular} \\ \hline
NoSQL Document Databases &
  10 &
  \begin{tabular}[c]{@{}l@{}}2013{[}2{]}; 2015{[}2{]}; \\ 2016{[}3{]}; 2017{[}2{]}; 2019\end{tabular} \\ \hline
NoSQL key-value Databases &
  2 &
  2011; 2014 \\ \hline
NoSQL Column Databases &
  3 &
  2015; 2018 \\ \hline
\begin{tabular}[c]{@{}l@{}}Hadoop, MapReduce, HDFS, Hive, Spark, \\ Object Oriented, XML, RDBMS, \\ Not about Data Modeling\end{tabular} &
  71 &
  2009-2020 \\ \hline
\multicolumn{1}{|r|}{\textbf{Total excluded}} &
  97 &
   \\ \hline
NoSQL Graph Databases, exclusively &
  7 &
  \begin{tabular}[c]{@{}l@{}}2014{[}2{]}; 2015; \\ 2016{[}2{]}; 2017{[}2{]}\end{tabular} \\ \hline
\begin{tabular}[c]{@{}l@{}}Multi Model NoSQL Databases, \\ including Graph\end{tabular} &
  9 &
  \begin{tabular}[c]{@{}l@{}}2013; 2015; 2016{[}2{]}; \\ 2017; 2018{[}3{]}; 2020\end{tabular} \\ \hline
\multicolumn{1}{|r|}{\textbf{Total included}} &
  16 &
   \\ \hline
\end{tabular}
\caption{Number of papers after inclusion and exclusion criteria}
\label{tab:incl-excl-criteria}
\end{table}

\clearpage

\begin{landscape}

\begin{table}[h]
\begin{tabular}{lllll}
\hline
\multicolumn{1}{|l|}{\textbf{\begin{tabular}[c]{@{}l@{}}Data Modeling \\ Characteristics\end{tabular}}} &
  \multicolumn{1}{c|}{{\cite{Kaur2013}}} &
  \multicolumn{1}{c|}{{\cite{DeVirgilio2014}}} &
  \multicolumn{1}{c|}{{\cite{Souza2014}}} &
  \multicolumn{1}{c|}{{\cite{Neumayr2015}}} \\ \hline
\multicolumn{1}{|l|}{\textbf{V-Dimensions}} &
  \multicolumn{1}{l|}{Not specified} &
  \multicolumn{1}{l|}{Not specified} &
  \multicolumn{1}{l|}{Not specified} &
  \multicolumn{1}{l|}{Not specified} \\ \hline
\multicolumn{1}{|l|}{\textbf{\begin{tabular}[c]{@{}l@{}}Data Models \\ (beside graph)\end{tabular}}} &
  \multicolumn{1}{l|}{Document} &
  \multicolumn{1}{l|}{None} &
  \multicolumn{1}{l|}{None} &
  \multicolumn{1}{l|}{None} \\ \hline
\multicolumn{1}{|l|}{\textbf{\begin{tabular}[c]{@{}l@{}}Graph Data \\ Models\end{tabular}}} &
  \multicolumn{1}{l|}{LPG} &
  \multicolumn{1}{l|}{Any (RDF, LPG, …)} &
  \multicolumn{1}{l|}{LPG} &
  \multicolumn{1}{l|}{RDF} \\ \hline
\multicolumn{1}{|l|}{\textbf{\begin{tabular}[c]{@{}l@{}}Models by \\ Abstraction \\ Level\end{tabular}}} &
  \multicolumn{1}{l|}{\begin{tabular}[c]{@{}l@{}}ERD - Conceptual\\ UML Class - Logical \\ MongoDB(BSON)-Physical\\ Graph - Logical\\ Neo4J(LPG) - Physical\end{tabular}} &
  \multicolumn{1}{l|}{\begin{tabular}[c]{@{}l@{}}ERD - Conceptual (or UML)\\ Oriented ER Diagram - Logical\\ Any NoSQL GraphDB - Physical\end{tabular}} &
  \multicolumn{1}{l|}{\begin{tabular}[c]{@{}l@{}}EB-ER - Conceptual \\ Graph Based - Logical\end{tabular}} &
  \multicolumn{1}{l|}{\begin{tabular}[c]{@{}l@{}}Domain Ontology - Conceptual \\ Aggregation Ontology - Conceptual \\ Named RDF-Logical\\ Jena TDB-Physical\end{tabular}} \\ \hline
\multicolumn{1}{|l|}{\textbf{\begin{tabular}[c]{@{}l@{}}Modeling \\ Languages\end{tabular}}} &
  \multicolumn{1}{l|}{UML, ER} &
  \multicolumn{1}{l|}{ER (or UML)} &
  \multicolumn{1}{l|}{ER} &
  \multicolumn{1}{l|}{OWL, SPARQL} \\ \hline
\multicolumn{1}{|l|}{\textbf{Methodology}} &
  \multicolumn{1}{l|}{data-driven} &
  \multicolumn{1}{l|}{model-driven} &
  \multicolumn{1}{l|}{model-driven} &
  \multicolumn{1}{l|}{ontology-driven} \\ \hline
\multicolumn{1}{|l|}{\textbf{Approach}} &
  \multicolumn{1}{l|}{forward} &
  \multicolumn{1}{l|}{forward} &
  \multicolumn{1}{l|}{forward} &
  \multicolumn{1}{l|}{forward and reverse} \\ \hline
\multicolumn{1}{|l|}{\textbf{\begin{tabular}[c]{@{}l@{}}Reverse \\ engineering \\ (source)\end{tabular}}} &
  \multicolumn{1}{l|}{Not aplicable} &
  \multicolumn{1}{l|}{Not aplicable} &
  \multicolumn{1}{l|}{Not aplicable} &
  \multicolumn{1}{l|}{instances and Aggregation Ontology} \\ \hline
 &
   &
   &
   &
   \\ \hline
\multicolumn{1}{|l|}{\textbf{\begin{tabular}[c]{@{}l@{}}Data Modeling \\ Characteristics\end{tabular}}} &
  \multicolumn{1}{c|}{{\cite{Priebe2015}}} &
  \multicolumn{1}{c|}{{\cite{Marcin2016}}} &
  \multicolumn{1}{c|}{{\cite{DeFreitas2016}}} &
  \multicolumn{1}{c|}{{\cite{Banerjee2016}}} \\ \hline
\multicolumn{1}{|l|}{\textbf{V-Dimensions}} &
  \multicolumn{1}{l|}{Variety; Volume;} &
  \multicolumn{1}{l|}{Not specified} &
  \multicolumn{1}{l|}{Variety} &
  \multicolumn{1}{l|}{Variety} \\ \hline
\multicolumn{1}{|l|}{\textbf{\begin{tabular}[c]{@{}l@{}}Data Models \\ (beside graph)\end{tabular}}} &
  \multicolumn{1}{l|}{\begin{tabular}[c]{@{}l@{}}Relational, Document, \\ File-System\end{tabular}} &
  \multicolumn{1}{l|}{None} &
  \multicolumn{1}{l|}{\begin{tabular}[c]{@{}l@{}}Relational, Key-Value, \\ Document, Column\end{tabular}} &
  \multicolumn{1}{l|}{Any (SQL and NoSQL)} \\ \hline
\multicolumn{1}{|l|}{\textbf{\begin{tabular}[c]{@{}l@{}}Graph Data \\ Models\end{tabular}}} &
  \multicolumn{1}{l|}{Any} &
  \multicolumn{1}{l|}{RDF} &
  \multicolumn{1}{l|}{LPG} &
  \multicolumn{1}{l|}{Any} \\ \hline
\multicolumn{1}{|l|}{\textbf{\begin{tabular}[c]{@{}l@{}}Models by \\ Abstraction \\ Level\end{tabular}}} &
  \multicolumn{1}{l|}{\begin{tabular}[c]{@{}l@{}}EER - Conceptual\\ BIM - Conceptual\\ Any industry reference \\         model DWH - Logical\\ Any - Physical\end{tabular}} &
  \multicolumn{1}{l|}{Native RDF Storage - Physical} &
  \multicolumn{1}{l|}{\begin{tabular}[c]{@{}l@{}}ERD - Conceptual\\ Relational - Logical\\ Any NoSQL - Logical\\ RDBMS metadata - Physical\end{tabular}} &
  \multicolumn{1}{l|}{\begin{tabular}[c]{@{}l@{}}Ontology - Conceptual\\ SQL and NoSQL - Physical\end{tabular}} \\ \hline
\multicolumn{1}{|l|}{\textbf{\begin{tabular}[c]{@{}l@{}}Modeling \\ Languages\end{tabular}}} &
  \multicolumn{1}{l|}{EER} &
  \multicolumn{1}{l|}{Not applicable} &
  \multicolumn{1}{l|}{ER} &
  \multicolumn{1}{l|}{OWL} \\ \hline
\multicolumn{1}{|l|}{\textbf{Methodology}} &
  \multicolumn{1}{l|}{model-driven} &
  \multicolumn{1}{l|}{data driven} &
  \multicolumn{1}{l|}{data-driven} &
  \multicolumn{1}{l|}{ontology driven} \\ \hline
\multicolumn{1}{|l|}{\textbf{Approach}} &
  \multicolumn{1}{l|}{forward} &
  \multicolumn{1}{l|}{Not applicable} &
  \multicolumn{1}{l|}{migration / interoperability} &
  \multicolumn{1}{l|}{forward} \\ \hline
\multicolumn{1}{|l|}{\textbf{\begin{tabular}[c]{@{}l@{}}Reverse \\ engineering \\ (source)\end{tabular}}} &
  \multicolumn{1}{l|}{Not aplicable} &
  \multicolumn{1}{l|}{Not applicable} &
  \multicolumn{1}{l|}{RDBMS metadata} &
  \multicolumn{1}{l|}{Not aplicable} \\ \hline
\end{tabular}
\caption{Data Modeling Characteristics}
\label{tab:dm1}
\end{table}

\begin{table}[h]
\begin{tabular}{lllll}
\hline
\multicolumn{1}{|l|}{\textbf{\begin{tabular}[c]{@{}l@{}}Data Modeling \\ Characteristics\end{tabular}}} &
  \multicolumn{1}{c|}{{\cite{Daniel2016}}} &
  \multicolumn{1}{c|}{{\cite{Comyn-Wattiau2017}}} &
  \multicolumn{1}{c|}{{\cite{Akoka2017}}} &
  \multicolumn{1}{c|}{{\cite{Abdelhedi2017}}} \\ \hline
\multicolumn{1}{|l|}{\textbf{V-Dimensions}} &
  \multicolumn{1}{l|}{Not mentioned} &
  \multicolumn{1}{l|}{Not specified} &
  \multicolumn{1}{l|}{\begin{tabular}[c]{@{}l@{}}Volume; Variety; \\ Velocity; Veracity\end{tabular}} &
  \multicolumn{1}{l|}{\begin{tabular}[c]{@{}l@{}}Velocity, Variety, \\ Volume\end{tabular}} \\ \hline
\multicolumn{1}{|l|}{\textbf{\begin{tabular}[c]{@{}l@{}}Data Models \\ (beside graph)\end{tabular}}} &
  \multicolumn{1}{l|}{None} &
  \multicolumn{1}{l|}{None} &
  \multicolumn{1}{l|}{None} &
  \multicolumn{1}{l|}{Column, Document} \\ \hline
\multicolumn{1}{|l|}{\textbf{\begin{tabular}[c]{@{}l@{}}Graph Data \\ Models\end{tabular}}} &
  \multicolumn{1}{l|}{LPG} &
  \multicolumn{1}{l|}{LPG} &
  \multicolumn{1}{l|}{LPG} &
  \multicolumn{1}{l|}{LPG} \\ \hline
\multicolumn{1}{|l|}{\textbf{\begin{tabular}[c]{@{}l@{}}Models by \\ Abstraction \\ Level\end{tabular}}} &
  \multicolumn{1}{l|}{\begin{tabular}[c]{@{}l@{}}UML Class Diagram -CIM\\ OCL Constraints -CIM\\ GraphDB metamodel-PIM\\ Blueprint API (LPG+Gremlin)-PSM\end{tabular}} &
  \multicolumn{1}{l|}{\begin{tabular}[c]{@{}l@{}}EER-CIM\\ GraphDB-PIM\\ Neo4J(LPG)-PSM\end{tabular}} &
  \multicolumn{1}{l|}{\begin{tabular}[c]{@{}l@{}}V’s EER-CIM\\ EER property graph - PIM\\ Neo4J(LPG)-PSM\end{tabular}} &
  \multicolumn{1}{l|}{\begin{tabular}[c]{@{}l@{}}UML Class - CIM\\ Generic Logical Model - PIM\\ Cassandra - PSM\\ MongoDB - PSM\\ Neo4J - PSM\end{tabular}} \\ \hline
\multicolumn{1}{|l|}{\textbf{\begin{tabular}[c]{@{}l@{}}Modeling \\ Languages\end{tabular}}} &
  \multicolumn{1}{l|}{\begin{tabular}[c]{@{}l@{}}UML, OCL, \\ ATL (aligned with QVT standard)\end{tabular}} &
  \multicolumn{1}{l|}{ER} &
  \multicolumn{1}{l|}{ER} &
  \multicolumn{1}{l|}{UML, Ecore, XML} \\ \hline
\multicolumn{1}{|l|}{\textbf{Methodology}} &
  \multicolumn{1}{l|}{model-driven} &
  \multicolumn{1}{l|}{model-driven} &
  \multicolumn{1}{l|}{model-driven} &
  \multicolumn{1}{l|}{model-driven} \\ \hline
\multicolumn{1}{|l|}{\textbf{Approach}} &
  \multicolumn{1}{l|}{forward} &
  \multicolumn{1}{l|}{reverse} &
  \multicolumn{1}{l|}{forward} &
  \multicolumn{1}{l|}{forward} \\ \hline
\multicolumn{1}{|l|}{\textbf{\begin{tabular}[c]{@{}l@{}}Reverse \\ engineering \\ (source)\end{tabular}}} &
  \multicolumn{1}{l|}{Not aplicable} &
  \multicolumn{1}{l|}{\begin{tabular}[c]{@{}l@{}}Metadata \\ (CREATE STATEMENTS)\\ Set of instances\end{tabular}} &
  \multicolumn{1}{l|}{Not applicable} &
  \multicolumn{1}{l|}{Not aplicable} \\ \hline
 &
   &
   &
   &
   \\ \hline
\multicolumn{1}{|l|}{\textbf{\begin{tabular}[c]{@{}l@{}}Data Modeling \\ Characteristics\end{tabular}}} &
  \multicolumn{1}{c|}{{\cite{Pulgatti2018}}} &
  \multicolumn{1}{c|}{{\cite{Zdepski2018}}} &
  \multicolumn{1}{c|}{{\cite{Bjeladinovic20181}}} &
  \multicolumn{1}{c|}{{\cite{Abdelhedi2020}}} \\ \hline
\multicolumn{1}{|l|}{\textbf{V-Dimensions}} &
  \multicolumn{1}{l|}{Variety} &
  \multicolumn{1}{l|}{Variety} &
  \multicolumn{1}{l|}{Variety} &
  \multicolumn{1}{l|}{Volume; Variety; Velocity;} \\ \hline
\multicolumn{1}{|l|}{\textbf{\begin{tabular}[c]{@{}l@{}}Data Models \\ (beside graph)\end{tabular}}} &
  \multicolumn{1}{l|}{\begin{tabular}[c]{@{}l@{}}Key-Value, Column, \\ Document\end{tabular}} &
  \multicolumn{1}{l|}{\begin{tabular}[c]{@{}l@{}}Relational, Key-Value, \\ Column, Document\end{tabular}} &
  \multicolumn{1}{l|}{Any (SQL and NoSQL)} &
  \multicolumn{1}{l|}{\begin{tabular}[c]{@{}l@{}}Key-Value, Column, \\ Document\end{tabular}} \\ \hline
\multicolumn{1}{|l|}{\textbf{\begin{tabular}[c]{@{}l@{}}Graph Data \\ Models\end{tabular}}} &
  \multicolumn{1}{l|}{LPG} &
  \multicolumn{1}{l|}{Any} &
  \multicolumn{1}{l|}{Any} &
  \multicolumn{1}{l|}{Any} \\ \hline
\multicolumn{1}{|l|}{\textbf{\begin{tabular}[c]{@{}l@{}}Models by \\ Abstraction \\ Level\end{tabular}}} &
  \multicolumn{1}{l|}{\begin{tabular}[c]{@{}l@{}}JSON - Logical\\ Any NoSQL - Physical\end{tabular}} &
  \multicolumn{1}{l|}{\begin{tabular}[c]{@{}l@{}}ER-Conceptual\\ All-Logical\end{tabular}} &
  \multicolumn{1}{l|}{\begin{tabular}[c]{@{}l@{}}Any - Conceptual\\ Relational - Logical\\ NoAM - Logical\\ Any RDBMS - Physical\\ Any NoSQL - Physical\end{tabular}} &
  \multicolumn{1}{l|}{\begin{tabular}[c]{@{}l@{}}UML Class Diagram - PIM\\ Cassandra - PSM\\ MongoDB - PSM\\ Neo4J - PSM\end{tabular}} \\ \hline
\multicolumn{1}{|l|}{\textbf{\begin{tabular}[c]{@{}l@{}}Modeling \\ Languages\end{tabular}}} &
  \multicolumn{1}{l|}{JSON} &
  \multicolumn{1}{l|}{ER} &
  \multicolumn{1}{l|}{Any} &
  \multicolumn{1}{l|}{UML, Ecore, XML, QVT} \\ \hline
\multicolumn{1}{|l|}{\textbf{Methodology}} &
  \multicolumn{1}{l|}{data-driven} &
  \multicolumn{1}{l|}{data-driven} &
  \multicolumn{1}{l|}{data-driven} &
  \multicolumn{1}{l|}{model-driven} \\ \hline
\multicolumn{1}{|l|}{\textbf{Approach}} &
  \multicolumn{1}{l|}{migration / interoperability} &
  \multicolumn{1}{l|}{forward} &
  \multicolumn{1}{l|}{forward, reverse} &
  \multicolumn{1}{l|}{reverse} \\ \hline
\multicolumn{1}{|l|}{\textbf{\begin{tabular}[c]{@{}l@{}}Reverse \\ engineering \\ (source)\end{tabular}}} &
  \multicolumn{1}{l|}{Not aplicable} &
  \multicolumn{1}{l|}{Not aplicable} &
  \multicolumn{1}{l|}{Metadata, Queries/Workload} &
  \multicolumn{1}{l|}{metadata of NoSQL Database} \\ \hline
\end{tabular}
\caption{Data Modeling Characteristics}
\label{tab:dm2}
\end{table}
\end{landscape}
\rmfamily 

\begin{received}
\end{received}

\end{document}